\documentclass[aip,reprint]{revtex4-1}
\usepackage{graphicx}
\usepackage[T1]{fontenc}
\usepackage{comment}

\draft 

\begin{document}


\title{Shaping Single Photons through Multimode Optical Fibers using Mechanical Perturbations} 



\author{Ronen Shekel}
\affiliation{Racah Institute of Physics, The Hebrew University of Jerusalem, Jerusalem 91904, Israel}

\author{Ohad Lib}
\affiliation{Racah Institute of Physics, The Hebrew University of Jerusalem, Jerusalem 91904, Israel}

\author{Rodrigo Gutiérrez-Cuevas}
\affiliation{{Institut Langevin, CNRS UMR 7587, ESPCI Paris, PSL Research University, 1 rue Jussieu, 75005 Paris, France}}

\author{Sébastien M. Popoff}
\affiliation{{Institut Langevin, CNRS UMR 7587, ESPCI Paris, PSL Research University, 1 rue Jussieu, 75005 Paris, France}}

\author{Alexander Ling}
\affiliation{{Centre for Quantum Technologies, 3 Science Drive 2, National University of Singapore 117543 Singapore}}
\affiliation{{Department of Physics, Faculty of Science, National University of Singapore, 2 Science Drive 3, 117551 Singapore}}

\author{Yaron Bromberg}
\email[]{yaron.bromberg@mail.huji.ac.il}
\affiliation{Racah Institute of Physics, The Hebrew University of Jerusalem, Jerusalem 91904, Israel}


\date{\today}

\begin{abstract}
The capacity of information delivered by single photons is boosted by encoding high-dimensional quantum dits in their transverse shape. Transporting such high-dimensional quantum dits in optical networks may be accomplished using multimode optical fibers, which support the low-loss transmission of multiple spatial modes over existing infrastructure. However, when photons propagate through a multimode fiber their transverse shape gets scrambled because of mode mixing and modal interference. This is usually corrected using free-space spatial light modulators, inhibiting a robust all-fiber operation. In this work, we demonstrate an all-fiber approach for controlling the shape of single photons and the spatial correlations between entangled photon pairs, using carefully controlled mechanical perturbations of the fiber. We optimize these perturbations to localize the spatial distribution of a single photon or the spatial correlations of photon pairs in a single spot, enhancing the signal in the optimized spot by over an order of magnitude. Using the same approach we show a similar enhancement for coupling light from a multimode fiber into a single-mode fiber. 
\end{abstract}


\maketitle 

\section{Introduction}
Quantum technologies are revolutionizing the fields of communication \cite{sidhu2021advances}, sensing \cite{polino2020photonic}, and computing \cite{nielsen2002quantum}. Due to their low decoherence, photons are excellent candidates for transmitting quantum information between distant points and are often referred to as "flying qubits" \cite{divincenzo2000physical,flamini2018photonic}. Photons also enjoy the advantage of existing mature optical fiber infrastructure offering low-loss transmission. In particular, multimode fibers (MMFs) can greatly boost the information capacity for both classical \cite{puttnam2021space, science2023topological} and quantum \cite{valencia2020unscrambling, Amitonova2020QKD_MMF,zhou2021high} channels, by encoding high-dimensional dits in the transverse shape of the photons. In the quantum regime, single photons have been delivered through MMFs for high-dimensional quantum key distribution (QKD) \cite{ding2017high}, and to increase the collection efficiency of free-space QKD \cite{jin2018demonstration}. In addition, photon pairs have been sent through the same MMF for creating programmable and high-dimensional quantum circuits \cite{defienne2016two,leedumrongwatthanakun2020programmable, makowski2023large}.

However, when classical or quantum light propagates through a complex medium such as an MMF, the information it carries is scrambled due to the random interference between the different modes of the MMF. In particular, the spatial distribution of a single photon in a superposition of multiple fiber modes exhibits a speckle pattern at the output facet, exactly like classical coherent light. When pairs of spatially entangled photons propagate in MMFs, the spatial distribution of the intensity at the output of the fiber is homogeneous, yet the two-photon spatial correlations exhibit a speckle pattern, coined two-photon speckle \cite{beenakker2009two, peeters2010observation}.

The speckle and two-photon speckle patterns can be refocused to a spot by tailoring the wavefront at the input of the fiber using a spatial light modulator (SLM) \cite{carpenter2014110x110, ploschner2015seeing, defienne2014nonclassical, pinkse2016programmable, Peng2018manipulation, defienne2018adaptive, lib2020real, cao2023controlling}. The wavefront can be found by measuring the transmission matrix of the fiber, from which the desired input is calculated \cite{popoff2010measuring}, or by using a "blind" optimization scheme to maximize the signal at the focal spot \cite{vellekoop2007focusing}. Both methods rely on manipulating light outside the fiber, requiring careful alignment of the shaped wavefront into the fiber, thus inhibiting a robust all-fiber operation. 
 
\begin{figure}[tbp]
\includegraphics[width=\linewidth]{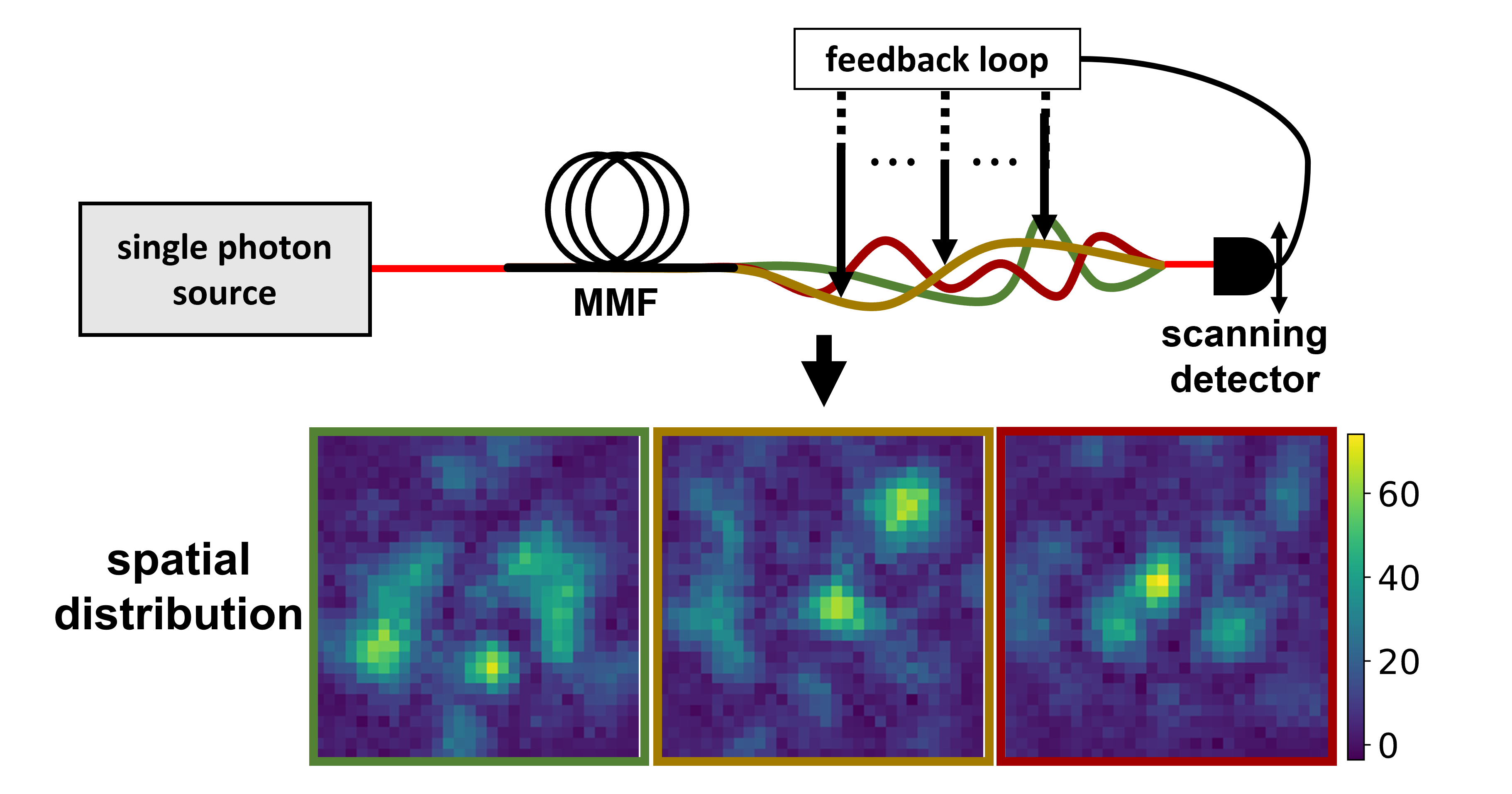}
\caption{\label{fig:concept} When a photon propagates through a multi-mode fiber, its spatial distribution is scrambled, producing a speckle pattern. This speckle is highly sensitive to mechanical perturbations of the fiber, so different mechanical fiber conformations produce different output patterns. By optimizing controlled mechanical perturbations of the fiber, we localize the spatial distribution at the output.}
\end{figure}

Recently, we demonstrated an all-fiber approach for shaping classical light through MMFs. The transmission matrix of an MMF is highly sensitive to mechanical perturbations of the fiber \cite{matthes2021learning}, as depicted in Fig.~\ref{fig:concept}. Turning this nuisance into a feature, we have recently shown that by applying carefully controlled mechanical perturbations on the fiber, we could shape classical light in both the spatial and spectral domains \cite{resisi2020wavefront, finkelstein2022spectral}. We refer to this system of mechanical perturbations along the fiber as the \textit{fiber piano}. 

In this work, we use the fiber piano to refocus the spatial distribution of a single photon and the spatial correlations of photon pairs that propagate through an MMF. In both cases, we achieve over an order of magnitude enhancement of the signal in the focal spot. We further demonstrate the ability to enhance the coupling of a single photon from an MMF into a single-mode fiber (SMF), which could potentially increase the collection efficiency in free-space quantum communications \cite{jin2018demonstration, cao2020distribution}. These results pave the way toward an all-fiber control of heralded single photons and photon pairs through MMFs.

\section{Results}
\begin{figure}[tbp]
\includegraphics[width=\linewidth]{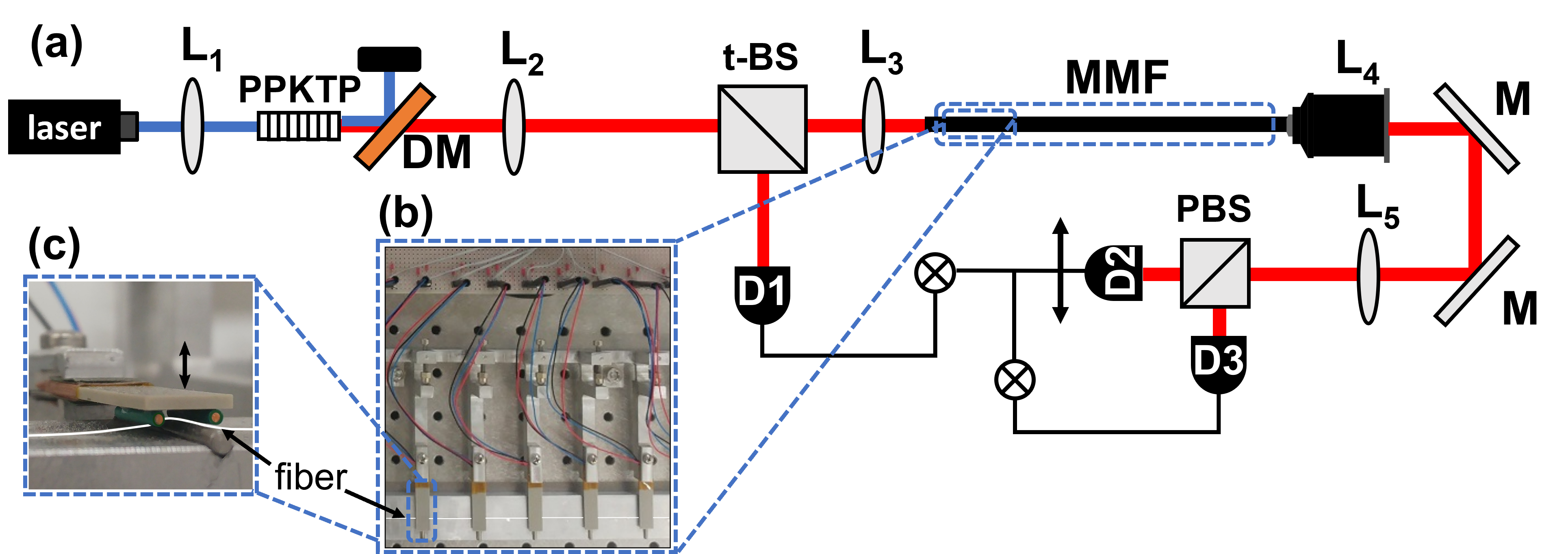}
\caption{\label{fig:setup} Experimental setup. A continuous-wave pump laser ($\lambda_p=403.8$nm) is focused using a lens ($L_1$) upon a $4$mm  type-0 PPKTP crystal, generating vertically polarized photon pairs via SPDC. The pump beam is discarded using a dichroic mirror (DM), and the photon pairs are imaged using lenses $L_2$ and $L_3$ to the input facet of a graded index MMF. Along the MMF 37 piezoelectric actuators are placed (b), inducing 3-point local bends on the fiber (c). The output facet of the MMF is then imaged using lenses $L_4$ and $L_5$ to the coincidence measurement plane, where there are two single-photon detectors. The photons are split using a polarizing beam splitter (PBS), so one photon reaches a fixed detector ($D_3$) and the other a scanning detector ($D_2$). In the heralded single-photon configuration, we place before the fiber a beam splitter with a tunable reflectivity, implemented by a $\frac{\lambda}{2}$ waveplate and a PBS. The PBS probabilistically routes one photon to the heralding detector ($D_1$) and one photon into the fiber. (M - mirror; t-BS - tunable beam splitter).}
\end{figure}

To demonstrate our all-fiber approach for shaping photons, we place 37 piezoelectric actuators along a graded index MMF with a diameter of $50\mu m$. Each actuator creates a three-point contact that induces a local bend, as depicted in Fig \ref{fig:setup}. We couple into the MMF photons generated by type-0 spontaneous parametric down-conversion (SPDC) (see methods). To study the shaping of a single photon passing through the MMF, we utilize the SPDC light to generate heralded single photons: we use a tunable beamsplitter in front of the MMF to probabilistically route one of the photons to a detector (D1 in Fig. \ref{fig:setup}a), and its twin photon to the MMF. We refer to this configuration as the single photon configuration. The image plane of the output facet of the MMF is then scanned by a second detector (D2). Coincidence events between the two detectors, i.e., simultaneous detection of a photon by each of the detectors, reveal the spatial distribution of the heralded photon that passes through the MMF. 

To study the correlations between pairs of photons that propagate together through the MMF, we tune the beamsplitter before the fiber to couple both photons to the MMF. We refer to this configuration as the two-photon configuration. We place two detectors (D2 and D3) at the image plane of the MMF output facet. One detector is fixed in place while the other scans the transverse plane, and the coincidence map reveals the two-photon spatial correlations. In both configurations, the collection of light to the detectors is done using fibers with a diameter of $50 \mu m$, which is slightly smaller than the size of a single speckle grain. Narrowband filters and a polarizing beam splitter are placed before the detectors to ensure the detection of approximately a single polarization and spectral mode. For full details, see Methods and Supplementary information. 

\begin{figure*}[htbp]
\includegraphics [width=\linewidth]{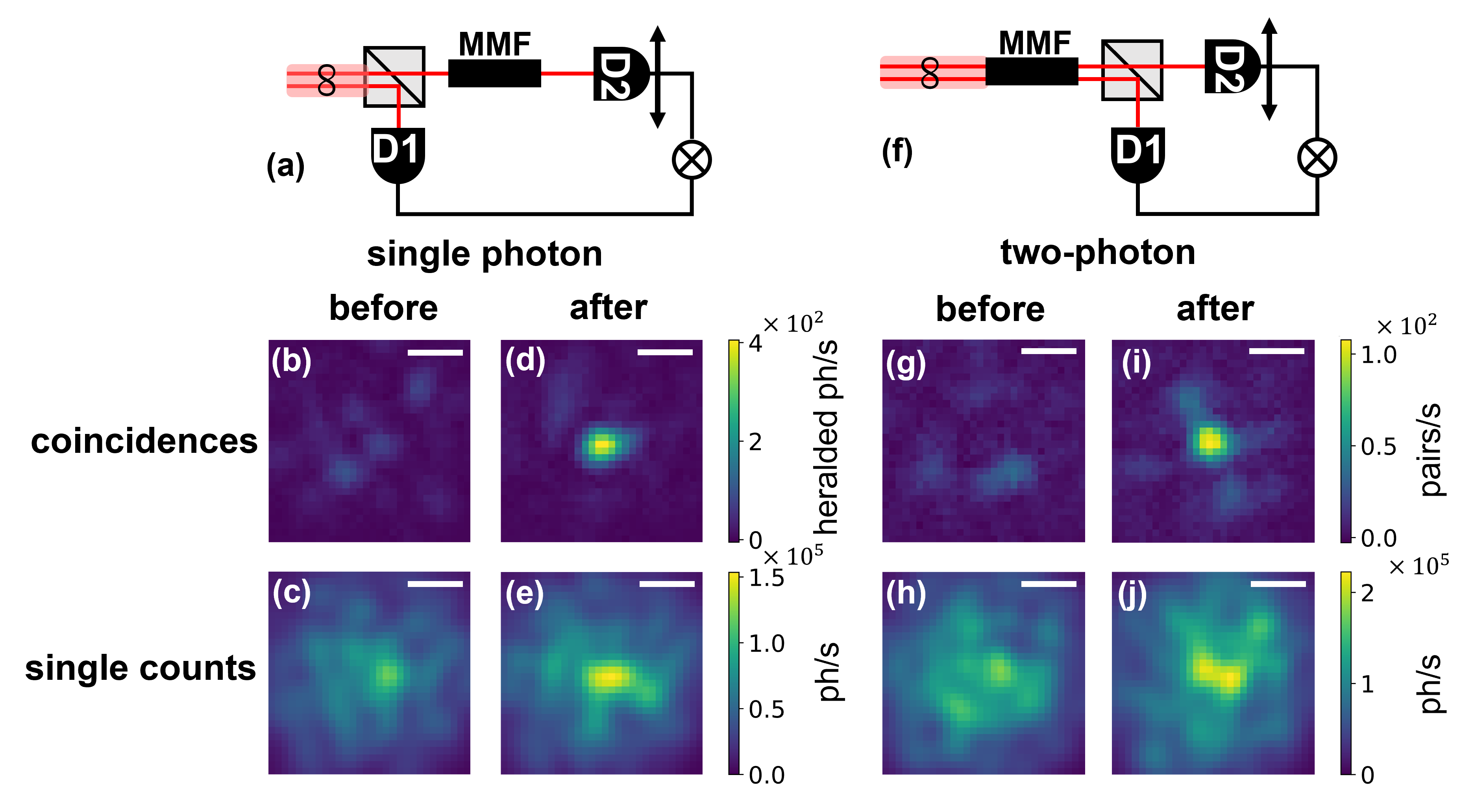}
\caption{\label{fig:optimization} We optimize the coincidence counts of single heralded photons (a-e) and photon pairs (f-j) passing through an MMF. Before the optimization process, the coincidence maps exhibit a speckle pattern (b,g). After the optimization, the coincidence maps are strongly localized (d, i), with an enhancement factor of $16\pm3$ and $12\pm2$ for the one photon and two-photon configurations, respectively. As expected, the single counts (c, e, h, j) exhibit a low contrast and show only a slight enhancement compared with the coincidences. The collection of light to the detectors is done using fibers with a diameter of $50 \mu m$, which is slightly smaller than the size of a single speckle grain. All scale bars represent $200\mu m$. (ph - photons).} \end{figure*}

As depicted in Fig. \ref{fig:optimization}, when either a single heralded photon (a) or photon pairs (f) propagate through the MMF, the coincidence maps exhibit a speckle pattern (b, g). The single counts, which are equivalent to the classical intensity, also exhibit a speckle pattern, but with a low contrast (c, h). The reason for the low contrast in the single counts is that the two-photon quantum state we generate is in a superposition of $\approx 15$ spatial modes that sum incoherently \cite{peeters2010observation}. For a detailed discussion on the contrast of the single counts see Supplementary information. 

To refocus the spatial distribution of the heralded single photons and the spatial correlations of the photon pairs, we use a particle swarm optimization (PSO) algorithm to find a set of mechanical perturbations that maximizes the coincidence rate of the two detectors. We define the enhancement as the ratio between the peak coincidence rate after optimization to the disorder average of the coincidence counts rate at the same point without optimization. For the single photon experiment we obtain an enhancement factor of $16\pm3$, and for the two-photon experiment we get an enhancement of $12\pm2$, as depicted in Fig. \ref{fig:optimization}d, i. 

We note that optimization of the coincidence rate at the target position slightly enhances also the single counts at that position (\ref{fig:optimization}e, j), yet by a factor ten times smaller. The small enhancement in the single counts shows that most of the enhancement of the correlations between the photons stems from unscrambling of the two-photon speckle pattern. The reason for this is that the two-photon signal is coherent, allowing to set the phases of different modes to interfere constructively at a desired spot, and is therefore easier to enhance by wavefront shaping than incoherent signals, similar to classical wavefront shaping. A detailed discussion on the contributions of the single counts, the polarization and the loss to the enhancement is given in the Supplementary information.

\begin{figure}[htbp]
\includegraphics[width=\linewidth]{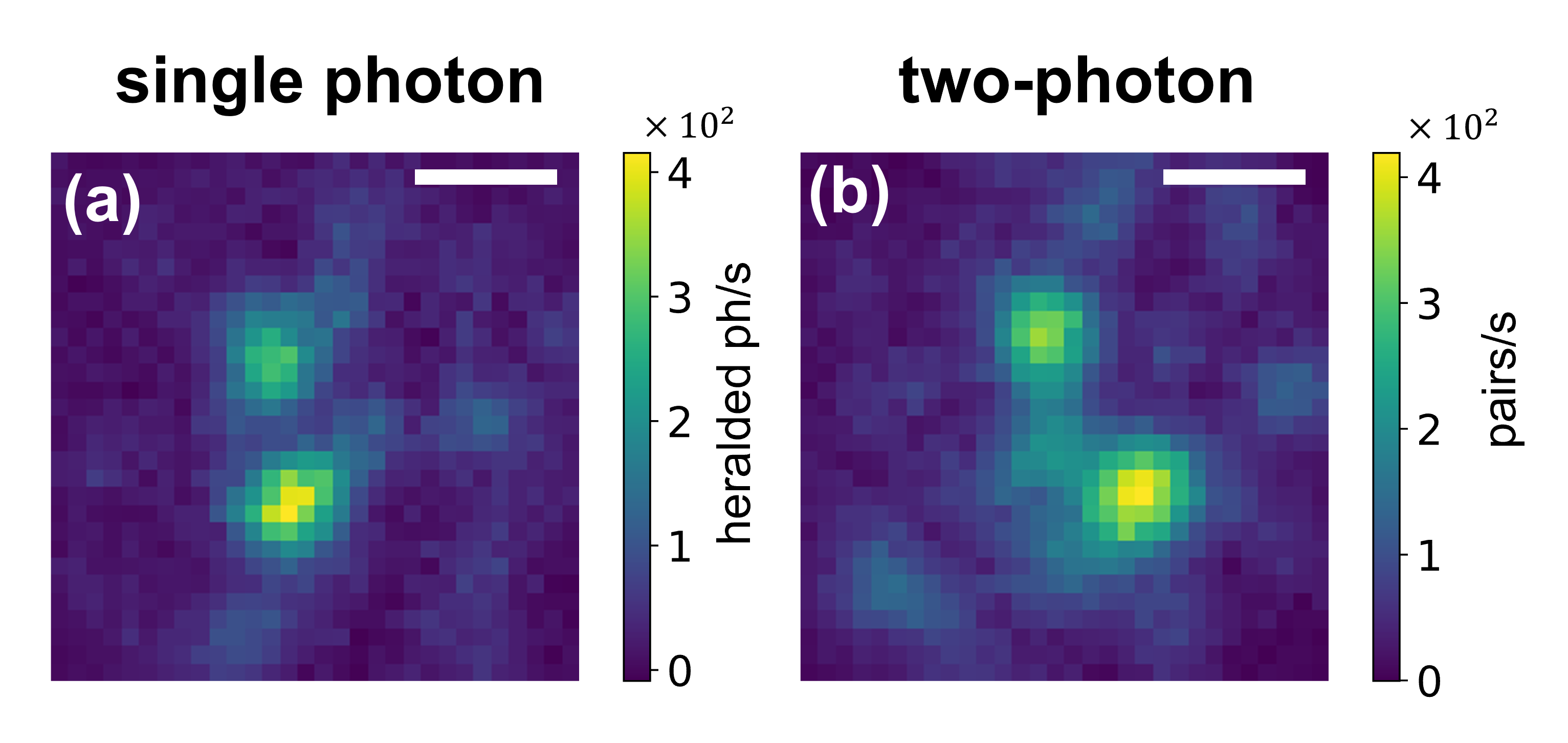}
\caption{\label{fig:two_spots} To demonstrate a higher degree of control using mechanical perturbations, we change the cost function to optimize the coincidence counts between the fixed detector and each of two spots. This is done both in the single photon configuration (a) and in the two-photon configuration (b). The collection of light to the detectors is done using fibers with a diameter of $100 \mu m$, which is on the order of a single speckle grain. The two scale bars represent $200\mu m$. (ph - photons).}
\end{figure}

Next, to demonstrate the flexibility of our approach to tailoring the shape of single photons and two-photon correlations, we perform the optimization to focus the coincidence maps on two spots. We optimize the coincidence counts between the fixed detector and each one of the two spots, simultaneously (see methods). Relative to the average coincidence counts at each spot, we achieve an enhancement of $4.1\pm0.4$ ($4.7\pm0.5$) on the upper spot and $5.7\pm0.5$ ($6.4\pm0.8$) on the lower spot in the single photon (two-photon) setting. The coincidence maps after the optimization process are shown in Fig. \ref{fig:two_spots}.

Finally, we consider the coupling of single photons from an MMF to an SMF. While sending single photons through MMFs is appealing when the information is encoded in the spatial degree of freedom, SMFs are more attractive in the case of temporal and polarization encoding. This is due to their lower dispersion and polarization mixing, better interface with superconducting detectors, and better components and infrastructure. However, in applications such as free-space quantum communications\cite{bedington2017progress,liao2017satellite}, coupling into SMFs limits the link efficiency, due to the need to map the distorted incoming wavefront to the single mode of the fiber. To demonstrate funneling of single photons through the MMF and coupling into an SMF, we couple heralded single photons into the MMF, and then optimize the coupling efficiency into a single polarization mode of an SMF using mechanical perturbations. We show that after optimization the collection efficiency into the SMF is enhanced by a factor of $11.3\pm0.6$. We show the optimization process for the coupling into the SMF in the Supplementary information. 


\section{Discussion}
We demonstrate the shaping of heralded single photons and photon pairs through MMFs via controlled mechanical perturbations, enhancing the signal at a focal spot by over an order of magnitude. In comparison to the standard approach of shaping the photons using SLMs, our approach is all-fiber, which can potentially offer a low-cost and robust solution. Most importantly, all-fiber shaping eliminates the need to achieve precise coupling of a large number of modes encoding the information of the quantum state into an MMF, which is particularly challenging in the high-dimensional case. 

Building on our demonstration of all-fiber shaping, we believe that utilizing recent demonstrations of the generation and detection of spatially entangled photons directly inside MMFs \cite{sulimany2022sourceandsorter, garay2023fiber} could enable all-fiber implementations of high-dimensional quantum technologies. For example, control of single photon propagation through MMFs could enable all-fiber high-dimensional QKD and entanglement distribution \cite{valencia2020unscrambling, goel2022inversedesign}, while control of photon pairs propagating through the same MMF could enable all-fiber linear quantum networks \cite{leedumrongwatthanakun2020programmable}. 

Another advantage of the fiber piano is that it allows tailoring of the entire transmission matrix of an MMF that supports $N$ guided modes, by increasing the number of controlled actuators in the experiment to approximately $N^2$. This is in contrast to the degree of control that can be achieved with an SLM in front of the fiber, which can only control $2N$ degrees of freedom, namely the complex amplitudes of the excited guided modes. Therefore, we believe that scaling up our proof-of-concept demonstration will not only yield larger enhancements \cite{resisi2020wavefront}, but also provide an all-fiber approach for programming arbitrary transformations on multiple input modes. This is analogous to the advantage of using multiplane light converters (MPLCs) with respect to a single SLM in both classical \cite{Treps2010progammabaleUnitary, labroille2014efficient, fontaine2019laguerre, butaite2022build} and quantum \cite{Fickler2020quantumgates, Fickler2021twophotoninterface, lib2022processing} experiments. We thus expect that while in this work our objective was to restore the correlations between specific modes, increasing the number of controlled actuators will allow for future applications in high-dimensional entanglement certification through MMFs \cite{valencia2020unscrambling, defienne2016two, goel2022inversedesign} and the realization of programmable linear quantum networks \cite{leedumrongwatthanakun2020programmable}.

Finally, similarly to adaptive optics with deformable mirrors which also relies on mechanical perturbations, we expect the optimization rate of our approach to be significantly improved by using a beacon laser for optimization \cite{defienne2016two, defienne2018adaptive}, fast electronics, and faster actuators, as discussed in the Supplementary information. We thus believe that our demonstration could be relevant to improving the link efficiency in free-space quantum communications through turbulence \cite{zheng2016free, ghalaii2022quantum}. This could be achieved by first coupling the distorted single photons into an MMF and then using mechanical perturbations to counter-effect the scattering in both the link and the MMF, yielding efficient coupling of the photons into an SMF.

\section{Methods} \label{Methods}
The experimental setup is presented in Fig \ref{fig:setup}. An $L=4$mm long type-0 PPKTP crystal is pumped by a continuous-wave laser (140 mW, $\lambda_p$ = 403.8 nm). The wavefront of the pump beam is focused on the crystal using a lens with a focal length of $L_1=200$mm. The pump profile in the crystal plane is approximately Gaussian with a waist of $w_0\approx 110\mu$m. Pairs of vertically polarized photons centered at $\lambda\approx 807.6nm$ are generated by the SPDC process. Since the divergence of the pump beam ($\lambda_p/w_0$) is much smaller than the divergence of the emitted pairs ($\sqrt{\lambda/L}$), the pairs are generated in an entangled superposition of multiple transverse modes \cite{walborn2010spatial}. The pump beam is separated from the entangled photons using a dichroic mirror and sent to a beam dump. The entangled photons are then imaged using two lenses with focal lengths $L_2=200$mm and $L_3=60$mm to the facet of an $\approx 1.8$m long multimode bare fiber (Corning ClearCurve OM2, graded index, numerical aperture of $0.2$ and diameter of $50 \mu m$). Between these lenses, the photons propagate through a $\lambda / 2$ waveplate and a polarizing beam splitter (PBS), used for switching between the single photon and the two-photon configurations.

Along the fiber we place 37 piezoelectric actuators (CTS NAC2225-A01), with a 3cm separation between adjacent actuators, creating three-point contacts that induce local bends along the fiber. The range of curvatures induced by the bends was chosen such that they would induce a significant change in the speckle pattern while causing a relatively low loss of a few percent per actuator. 

The far end of the fiber is imaged using a 20X objective lens (Olympus RMS20X) and another lens with focal length $L_5=300$mm, through another PBS, to the coincidence measurement plane. The coincidence measurements are performed after passing through $3$nm FWHM bandpass filters (Semrock LL01-808-12.5), using $50\mu$m multimode fibers (Thorlabs FG050LGA, step index, numerical aperture of $0.22$) coupled to single photon detectors (Excelitas SPCM-AQRH). The coincidence counts are measured using a time tagger (Swabian, Time tagger 20), with a coincidence window of $1$ns. All coincidence counts shown in this paper are after subtracting accidental counts. 

In the single photon configuration, the heralding photon is collected between L2 and L3, using a 50$\mu m$ multimode fiber (Thorlabs FG050LGA). When we optimize for two spots we use two fibers out of a fiber bundle with three 100$\mu$m diameter cores, with 200$\mu$m separation between core centers (FiberTech Optica FTO-CTFOLA). In the two-photon configuration when optimizing for two spots, we used also for the second fiber a $100\mu$m multimode fiber. To enhance coincidence rates at the two spots, $c_1$ and $c_2$, such that $c_1 \approx c_2$, we maximize the cost function $\sqrt{c_1} + \sqrt{c_2} - \alpha\cdot|c_1 - c_2|$, where $\alpha\approx0.04$ was adjusted to balance the tradeoff between a high signal and an equal signal at both spots.  

When coupling heralded single photons from an MMF into an SMF, we connect the fibers using a standard connector, and pass through an inline fiber polarizer (Thorlabs ILP780PM-FC) before arriving at the detectors. 

\section*{Funding Information}
This research was supported by the Zuckerman STEM Leadership Program and the ISF-NRF Singapore joint research program (Grant No. 3538/20).

\section*{Disclosures}
The authors declare no conflicts of interest.

\section*{Data availability}
The data that support the findings of this study are available from the corresponding author upon reasonable request.

\bibliography{Main}

\end{document}



\title{Shaping Single Photons through Multimode Optical Fibers using Mechanical Perturbations - Supplementary Information} 



\author{Ronen Shekel}
\affiliation{Racah Institute of Physics, The Hebrew University of Jerusalem, Jerusalem 91904, Israel}

\author{Ohad Lib}
\affiliation{Racah Institute of Physics, The Hebrew University of Jerusalem, Jerusalem 91904, Israel}

\author{Rodrigo Gutiérrez-Cuevas}
\affiliation{{Institut Langevin, CNRS UMR 7587, ESPCI Paris, PSL Research University, 1 rue Jussieu, 75005 Paris, France}}

\author{Sébastien M. Popoff}
\affiliation{{Institut Langevin, CNRS UMR 7587, ESPCI Paris, PSL Research University, 1 rue Jussieu, 75005 Paris, France}}

\author{Alexander Ling}
\affiliation{{Centre for Quantum Technologies, 3 Science Drive 2, National University of Singapore 117543 Singapore}}
\affiliation{{Department of Physics, Faculty of Science, National University of Singapore, 2 Science Drive 3, 117551 Singapore}}

\author{Yaron Bromberg}
\email[]{yaron.bromberg@mail.huji.ac.il}
\affiliation{Racah Institute of Physics, The Hebrew University of Jerusalem, Jerusalem 91904, Israel}


\date{\today}

\pacs{}

\maketitle 

\makeatletter
\renewcommand \thesection{S\@arabic\c@section}
\renewcommand\thetable{S\@arabic\c@table}
\renewcommand \thefigure{S\@arabic\c@figure}
\makeatother

\section{Mode number estimation} \label{supp:Mode_Estimation}
\subsection{Spatial modes - the Schmidt number}
The Schmidt number $K$ quantifies the entanglement of a quantum state. In the case of a maximally entangled state in the spatial domain, it is related to the number of different incoherent spatial modes that a single photon occupies when tracing out its twin photon. After propagating through the MMF, each of these modes produces a different speckle pattern. When looking at the single counts, these modes are summed incoherently, reducing the image contrast to zero for high Schmidt numbers. For a finite number of $K$ modes, we can expect to see a speckle pattern with a reduced contrast $C$ which is given by $C=\frac{1}{\sqrt{K}}$ \cite{goodman2007speckle}. 

To get a good coupling efficiency of the SPDC light into the multimode fiber, the Schmidt number must be lower than the $\approx300$ modes per polarization that the fiber supports. This guides us in choosing the parameters of the pump beam on the crystal, and here we quantify the Schmidt number of the state created by the SPDC process. 

To estimate the Schmidt number we measure the single counts at a given spot at the output of the fiber, for $\approx1900$ different random actuator configurations. We calculate the standard deviation of the measured single counts, which is equivalent to the contrast $C$ of the speckle pattern at the measured spot. From the contrast we compute the Schmidt number $K=\frac{1}{C^2}$. However, besides the spatial modes, there are other factors that can reduce contrast, such as the photons' bandwidth, the spatial resolution in the acquisition, and polarization effects. Assuming these factors affect the contrasts of the single counts and coincidence in the same manner, the ratio between the number of estimated modes in the single counts and in the coincidence counts can help us get a better estimate of the number of spatial modes. This is due to the fact that for maximally entangled states the contrast of the two-photon speckle pattern is $1$ \cite{beenakker2009two}. 

We calculate $\approx1.2$ modes in the coincidence counts, and arrive at an estimate of $K\approx15$ SPDC spatial modes. 

\subsection{Polarization and spectral modes}
Using a PBS before the detectors ensures we choose a single polarization mode, where one photon is horizontally polarized, and the other is vertically polarized. This has the same effect as inserting a polarizer and then a beam splitter before the detectors, as it selects a single polarization mode. 

A measure for the number of spectral modes in a given fiber is the spectral correlation width $\Delta\lambda$, defined as the wavelength shift required for creating two uncorrelated speckle patterns at the output of the fiber\cite{Redding:13}. To measure $\Delta\lambda$, we send light from a superluminescent diode (spectral bandwidth of $\approx14$nm) through the MMF, and collect at the output light from a single speckle grain into a spectrometer. The variation in intensity for different wavelengths originates from the spectral correlation width, which we quantify by the full-width-half-max of the autocorrelation of the measured spectrum. We estimate $\Delta\lambda\approx2.8$nm, which is comparable to the width of the bandpass filters we use for the SPDC light. We therefore conclude that the spectral correlation width of the fiber has a minor effect on the contrast of the two-photon speckle that we measure and on the enhancements that we obtain. 

\section{The optimization process} \label{supp:optimization}

\subsection{Contributions to the enhancement}
When optimizing the correlations, we showed an enhancement of over an order of magnitude. There are several mechanisms by which mechanical perturbation can contribute to the enhancement, and here we quantify them. Two global mechanisms that might contribute to the enhancement are a global polarization rotation to the selected polarization mode and a global reduction of the loss induced by the actuators. These effects enhance the total signal at the output of the fiber but do not change the shape of the pattern. We quantify these effects using a camera before and after the optimization process and find that these factors change by less than $12\%$. 

Another mechanism that can enhance the signal at the focused spot is amplifying the single counts. The single counts are indeed enhanced by a factor of $1.7$ ($1.5$) in the single photon (two-photon) configuration. This slight enhancement is expected due to the incoherent sum of different SPDC modes, one of which we enhance.

The main contribution to the enhancement comes from unscrambling the two-photon speckle by creating constructive interference at the focused spot. To quantify this, we normalize the enhancement by the total coincidence counts before and after the optimization. The normalized enhancement we obtain is $12\pm2$ ($7\pm1$) for the single photon (two-photon) configuration. The enhancement due to all of the other mechanisms mentioned above can be quantified by the ratio between the total coincidence counts after and before optimization, which is $1.4$ ($1.8$) for the single photon (two-photon) configuration.

\subsection{Optimizing the single counts does not work}
As mentioned above, propagation through an MMF of each SPDC spatial mode results in a different speckle pattern, and measuring the single counts will display an incoherent sum of these patterns. Conversely, measuring coincidence events effectively picks out a single mode and measures its spatial distribution. For this reason, trying to focus the single counts is equivalent to focusing all different modes simultaneously and would result in very low enhancement when looking at the coincidence counts for a specific mode. 

\begin{figure}[htbp]
\includegraphics[width=\linewidth]{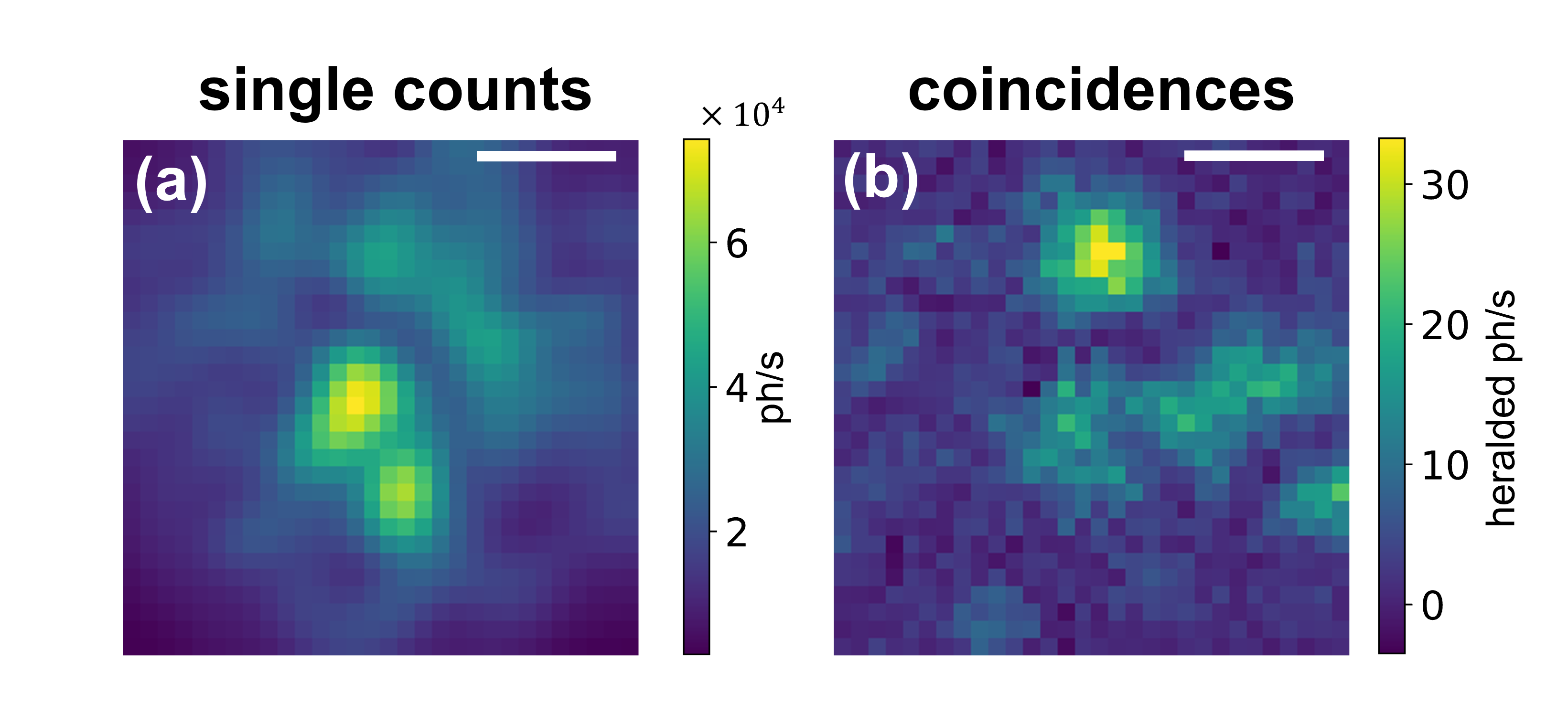}
\caption{\label{fig:singles_not_enough} We run the optimization process using the single counts as feedback. While the single counts are enhanced (a), this is not sufficient for enhancing the correlations (b). Both scale bars represent $200\mu m$. (ph - photons).}
\end{figure}

To demonstrate this, we run the optimization using the single counts for feedback. As depicted in Fig. \ref{fig:singles_not_enough}, we see that even though the single counts are slightly enhanced, the coincidence counts are not enhanced at all. 

\subsection{Optimization speed}
The optimization speed is determined by the signal acquisition time and by the speed of the electronics and mechanical actuators. Since we use the coincidence counts as our signal, our acquisition time for each iteration is $5$ seconds for the single photon configuration and $8$ seconds for the two-photon configuration. In addition, our current electronics and mechanical actuators allow changing configurations every $\approx 0.3$ seconds. This results in a slow optimization, which typically takes a few tens of minutes. However, using a beacon laser could reduce the acquisition time by many orders of magnitude \cite{defienne2016two, defienne2018adaptive}. In addition, faster electronics and mechanical actuators, such as those used in deformable mirrors, can result in rates approaching the KHz regime \cite{archer2016dynamic}, nearly three orders of magnitude faster than our current system. Furthermore, since our approach to modulating the transmission matrix of the fiber is general and not limited to mechanical perturbations, it could be directly transferred to other types of actuators, e.g., in-fiber acousto-optical or electro-optical modulators, which could reach even higher rates.

\subsection{Optimization process for coupling into an SMF}
\begin{figure}[htbp]
\includegraphics[width=\linewidth]{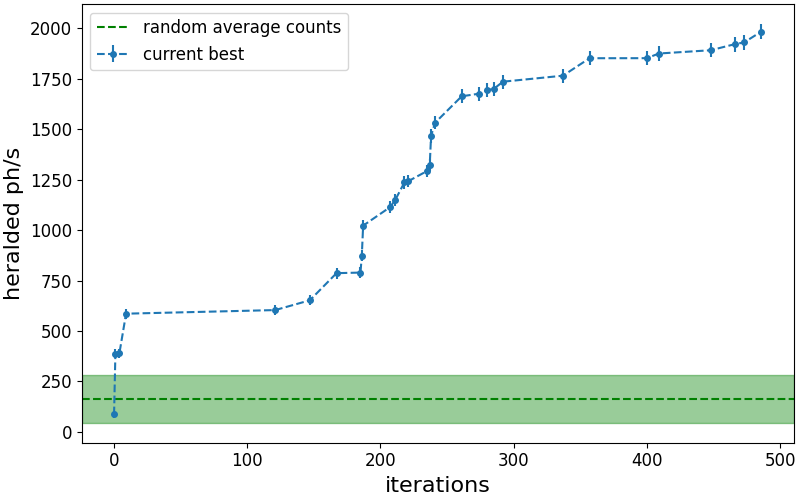}
\caption{\label{fig:SMF_optimization} We optimize the coupling of heralded single photons from an MMF into an SMF. The green dashed line and envelope correspond to the mean and standard deviation of the counts for the first $100$ configurations, which are random. Within a few hundred iterations we achieve an enhancement of over an order of magnitude.}
\end{figure}

We show the optimization process of coupling heralded single photons from an MMF into an SMF. As depicted in Fig. \ref{fig:SMF_optimization}, within a few hundred iterations (corresponding to $\approx 20$ minutes), the coupling into the SMF is enhanced by over an order of magnitude.

\bibliography{supplementary}